\documentclass[prd,showpacs,amsmath,amssymb,superscriptaddress,nofootinbib,showkeys]{revtex4}
\usepackage{amssymb}
\usepackage{amsmath}
\usepackage[cp1251]{inputenc}
\usepackage[T2A]{fontenc}
\usepackage{graphicx}
\usepackage{floatflt}

\DeclareGraphicsExtensions{.pdf,.eps,.png,.jpg}

\begin{document}
\title{Reconstruction method  in the kinetic
gravity braiding theory with shift-symmetric}

\author{Ruslan K. Muharlyamov}
\email{rmukhar@mail.ru} \affiliation{Department of General
Relativity and Gravitation, Institute of Physics, Kazan Federal
University, Kremlevskaya str. 18, Kazan 420008, Russia}

\author{Tatiana N. Pankratyeva}
\email{ghjkl.15@list.ru} \affiliation{Department of Higher
Mathematics, Kazan State Power Engineering University,
Krasnoselskaya str. 51, Kazan 420066, Russia}


\begin{abstract}
We present a reconstruction method for flat
Friedman-Robertson-Walker (FRW) spacetime in a subclass of
Horndeski theory -- specifically shiftsymmetric, the kinetic
gravity braiding (KGB) theory  with a non-vanishing conserved
current. Choosing the form of the Hubble parameter and kinetic
density $X$, we restore the functions $G_2$, $G_3$ of the KGB
model. In order to determine whether the model is free of ghosts
and Laplacian instabilities and thus cosmologically viable, two
conditions related to scalar perturbations are checked. Initially,
the Lagrangian does not include the term that describes, for
example, the perfect fluid with the EoS parameter $w\neq -1$. This
fluid can provide a dynamic solution $H(t)$, $X(t)$. In the
presented method, dynamic solutions are provided by a nonzero
scalar charge associated with the shift symmetry
$\phi\rightarrow\phi+\phi_0$. Reconstruction examples  are given
for models: an perfect fluid, a unified description dark
energy-dark matter, a post-inflationary transition to the
radiation-dominated phase.

\end{abstract}



\maketitle

\section{Introduction}

General relativity (GR) is confirmed by many observational data
\cite{Abbott1, Abbott2, Scientific, Abbott3, Collaboration}. But
to explain the accelerated expansion of the Universe and other
observational facts, within the framework of GR one has to involve
the hypothesis of the existence of an exotic substance called the
relativistic dark energy (DE). An alternative solution is to
construct modified gravity models. Also, an important motive for
modifying the gravity theory  is the impossibility of
extrapolating GR to all scales. There is a well-known problem of
compatibility of GR with quantum mechanics \cite{Deser, Hawking,
Martellini, Donoghue}.

The Horndeski gravity (HG) occupies a special place among the
modified models \cite{Horndeski}. The field equations in GR are
differential equations of the second order, thus, evading
Ostrogradski instabilities arising \cite{Ostrogradski,Woodard}.
The HG is the most general variant of the scalar-tensor theory of
gravitation with motion equations of the second order. The HG
includes, as special cases, all known examples of scalar-tensor
gravitational theories of the second order (Brans-Dicke theory,
$f(R)$ gravity, quintessence, etc.) \cite{Copeland,Joyce,
Armendariz-Picon, Tsujikawa, Deffayet, Kobayashi}. The action
density of HG contains several functions that provide a broad
phenomenology. This makes it possible to solve important
cosmological and astrophysical problems (screening of the
cosmological constant, kinetic inflation, late de Sitter stage,
hairy black holes, etc.) \cite{Sushkov0, Appleby, Sotiriou,
Babichev, Maselli}.

Various criteria for the selection of alternative gravity theories
and physical models proposed by them are discussed. In HG the
derivative self-couplings of the scalar field screen the
deviations from GR at small scales or high densities through the
Vainshtein mechanism \cite{Vainshtein}, thus satisfying solar
system and early universe constraints \cite{Chow, DeFelice}. After
the gravitational wave event GW170817 \cite{Collaboration}, the
modified theories of gravity, in which the propagation speed of
gravitational waves can differ from the speed of light, have a
dubious status \cite{Baker,Bettoni}. When considering modified
models, one has to worry about the presence of ghosts as well as
Laplacian instabilities \cite{DeFelice1,Appleby1}. In spite of
these constraints, the HG has retained most of the viable models.

The HG is determined by the following action density
\begin{eqnarray}\label{lagr1} L_H=\sqrt{-g}\Big(\mathcal{L}_2+\mathcal{L}_3+\mathcal{L}_4+\mathcal{L}_5\Big) \,,\end{eqnarray}
one has (in the parameterization of ref.\cite{Kobayashi1}):
$$\mathcal{L}_2 = G_2(\phi,X)\,,\, \mathcal{L}_3 = -
G_3(\phi,X)\Box\phi\,,$$
$$\mathcal{L}_4 = G_{4}(\phi,X) R +G_{4X}(\phi,X) \left[ (\square
\phi )^{2}-(\nabla_\mu \nabla_\nu \phi)^2  \right]\,,$$
\begin{equation} \mathcal{L}_5 = G_{5}(\phi,X) G_{\mu\nu}\,\nabla^\mu \nabla^\nu
\phi -\frac{1}{6}G_{5X}   \left[\left( \Box \phi \right)^3 -3 \Box
\phi (\nabla_\mu \nabla_\nu \phi)^2 + 2\left(\nabla_\mu \nabla_\nu
\phi \right)^3 \right], \label{lagr2}
\end{equation}
respectively, where $g$ is the determinant of metric tensor
$g_{\mu\nu}$; $R$ is the Ricci scalar and $G_{\mu\nu}$ is the
Einstein tensor; the factors $G_{i}$ ($i=2,3,4,5$) are arbitrary
functions of the scalar field $\phi$ and the canonical kinetic
term, $X=-\frac{1}{2}\nabla^\mu\phi \nabla_\mu\phi$. We consider
the definitions $G_{iX}\equiv \partial G_i/\partial X$,
 $(\nabla_\mu \nabla_\nu \phi)^2\equiv\nabla_\mu \nabla_\nu \phi
\,\nabla^\nu \nabla^\mu \phi$, and $\left(\nabla_\mu \nabla_\nu
\phi \right)^3\equiv \nabla_\mu \nabla_\nu \phi  \,\nabla^\nu
\nabla^\rho \phi \, \nabla_\rho \phi  \nabla^\mu \phi$.

We will consider the KGB theory: $G_2\neq 0$, $G_3\neq 0$,
$G_4=1/(16\pi)$, $G_5=0$. The shift-symmetric sector of the theory
corresponds to the choice $G_i=G_i(X)$. Then the action density
(\ref{lagr1}) takes the form
\begin{eqnarray}\label{lagr3} L_H=\sqrt{-g}\left(\frac{R}{16\pi}+G_2(X)-G_3(X)\Box\phi\right) \,.\end{eqnarray}
This theory is not sensitive to the constraints associated with
the gravitational event GW170817. The KGB cosmologies have been
extensively studied in the past, for example in \cite{Damien}. In
these models, the Universe smoothly evolves from contraction to
expansion, suffering neither from ghosts nor Laplacian
instabilities around the turning point. The end-point of the
evolution can be a standard radiation-domination era or an
inflationary phase. The necessary constraints for Lagrangians are
formulated, which  are necessary to obtain a healthy bounce. In
the work \cite{Helpin},  authors analyze what happens when the
Horndeski Lagrangian is varied within the Palatini approach by
considering the metric and connection as independent variables.
For example, in the context of the connection with the Palatini
approach, it was shown that the KGB subset of the HG cosmologies
may not contain ghosts.

We present a reconstruction method for flat
Friedman-Robertson-Walker (FRW) spacetimes in the subclass of HG
(\ref{lagr3}) with a non-vanishing conserved current. Choosing the
form of the Hubble parameter $H$ and kinetic density $X$, we
restore the functions $G_2$ and $G_3$. Next, we check if the model
is free of ghosts and Laplacian instabilities and thus
cosmologically viable. This method allows for a realistic model of
the Universe to simply reconstruct some field theory. The work
\cite{Bernardo} considers reconstruction in the KGB theory with
the vanishing conserved current and the shiftsymmetric. In this
theory, the dynamics in time of parameters $H(t)$ and $X(t)$ is
provided by the presence of other matter with variable density
$\rho(t)$. We are considering a model that contains only a scalar
field $\phi$ as source. The dynamism of functions $H(t)$ and
$X(t)$ is provided by the \emph{non-vanishing} conserved current.
For example, we will reconstruct theories for models: an perfect
fluid, a unified description dark energy-dark matter, a
post-inflationary transition to the radiation-dominated phase.

\section{Field equations}\label{sec2}

We consider a spatially-flat FRW metric
\begin{equation}\label{metr}ds^2=-dt^2+a^2(t)(dx^2+dy^2+dz^2)\,.
\end{equation}
Denoting $H =\dot{a}/a$ the Hubble parameter, the non-trivial
gravitational equations for (\ref{lagr3}) are
\begin{equation}\label{g00}\frac{3H^2}{8\pi}=-G_2+\dot{\phi}^2G_{2X}+3H\dot{\phi}^3G_{3X} \equiv\rho_{\phi}\,,\end{equation}
\begin{equation}\label{g11}\frac{-2\dot{H}-3H^2}{8\pi}=G_2-G_{3X}\ddot{\phi}\dot{\phi}^2\equiv p_{\phi}\,,\end{equation}
where $\rho_{\phi}$ is the energy density  and $p_{\phi}$ is the
pressure of the scalar.

The scalar equation
\begin{equation}\label{conservcurrent}\nabla_\mu\left(-\nabla^\mu\phi[G_{2X}-G_{3X}\Box \phi\,]+G_{3X}\nabla^\mu X\right)=0\end{equation}
has the structure of current conservation, due to the theory
invariance under shifts $\phi\rightarrow\phi+\phi_0$. In the FRW
space, the scalar current
\begin{equation}\label{current}J^\mu\equiv-\nabla^\mu\phi[G_{2X}-G_{3X}\Box \phi\,]+G_{3X}\nabla^\mu X\end{equation}
reduces to
\begin{equation}\label{FRWcurrent}J^\mu=\delta^\mu_0\dot{\phi}(G_{2X}+3HG_{3X}\dot\phi)\,.\end{equation}
The scalar field equation (\ref{conservcurrent}) becomes
\begin{equation}\label{concurrentFRW}\frac{1}{a^3}\frac{d}{dt}\left[a^3\dot{\phi}(G_{2X}+3HG_{3X}\dot\phi)\right]=0.\end{equation}
It has the first integral
\begin{equation}\label{scalph}\dot{\phi}(G_{2X}+3HG_{3X}\dot\phi)=\frac{C_0}{\,a^3}\,,\end{equation}
where $C_0$ is the scalar charge associated with the shift
symmetry $\phi\rightarrow\phi+\phi_0$.  Among equations
(\ref{g00}), (\ref{g11}) and (\ref{scalph}) only two are
independent. We have four functions ($H(a)$, $X(a)$, $G_2$, $G_3$)
and two independent equations. Therefore, we have two degrees of
freedom. By setting the expansion scenario of the Universe $H(a)$
and the kinetic dependence $X(a)$, we find the functions $G_2(X)$
and $G_3(X)$. We will note an important property of the presented
model. If the right-hand side of equation (\ref{scalph}) is zero,
then the system (\ref{g00}), (\ref{g11}), (\ref{scalph}) can only
have a stationary solution $H,\, X=const$. To provide a broad
variety of cosmological scenarios, we will consider the model with
the non-vanishing scalar current, i.e $C_0\neq 0$ and
$\dot{\phi}\neq 0$.

Combining the equations (\ref{g00}), (\ref{g11}) and
(\ref{scalph}), we arrive at the equalities
\begin{equation}\label{recg2}G_2=-\frac{3H^2[a(X)]}{8\pi}+\frac{\varepsilon C_0\sqrt{2X}}{a^3(X)}\,,\end{equation}
\begin{equation}\label{recg23X}G_{3X}=\frac{H'_X}{4\pi\varepsilon\sqrt{2X}}+\frac{C_0a'_X}{Ha^4(X)}\,,\end{equation}
where $\varepsilon=\pm 1$ defines the sign of the derivative
$\dot\phi=\varepsilon\sqrt{2X}$. Here we took into account
$\dot{\phi}\ddot{\phi}=\dot{X}=aH/a'_X$. It is assumed that for
the given function $X(a)$ there is an inverse function $a(X)$.

It is clear that the Hubble parameter $H(a)$ is set from
cosmological considerations. What will determine the choice of
kinetic density $X(a)$? This arbitrariness will be used to ensure
the viability of the model. By choosing the function $X(a)$, we
will exclude ghosts and Laplacian instabilities in the region of
applicability of the $ H (a) $ model. Thus, two conditions related
to scalar perturbations must be satisfied \cite{DeFelice1}:
\begin{equation}c_S^2\equiv\frac{3(2w_1^2w_2H-w^2_2w_4+4w_1w_2\dot w_1-2w_1^2\dot w_2)}{w_1(4w_1w_3+9w_2^2)}\geq 0\,,\label{cS}
\end{equation}
for the avoidance of Laplacian instabilities associated with the
scalar field propagation speed, and
\begin{equation}
Q_{S}\equiv\frac{w_{1}(4w_{1}w_{3}+9w_{2}^{2})}{3w_{2}^{2}}>0\,,
\label{QS}
\end{equation}
for the absence of ghosts, where
\begin{eqnarray}
&&w_{1}  \equiv  2\,(G_{{4}}-2\,
XG_{{4X}})-2X\,(G_{{5X}}{\dot{\phi}}H-G_{{5\phi}}) \,,
\label{w1def}\\
&&w_{2}  \equiv  -2\, G_{{3X}}X\dot{\phi}+4\,
G_{{4}}H-16\,{X}^{2}G_{{4{ XX}}}H+4(\dot{\phi}G_{ {4\phi X}}-4H\,
G_{{4X}})X+2\, G_{{4\phi}}\dot{\phi} \nonumber
\\
 &  & \ \ \ \ \ \ \ \,
 +8\,{X}^{2}HG_{{5\phi X}}+2H\, X\,(6G_{{5\phi}}-5\,
G_{{5X}}\dot{\phi}{H})-4G_{{5{
XX}}}{\dot{\phi}}X^{2}{H}^{2}\,,\\
&& w_{3}  \equiv  3\, X(G_{2{X}}+2\, XG_{{
2XX}})+6X(3X\dot{\phi}HG_{{3{ XX}}}-G_{{3\phi X}}
X-G_{{3\phi}}+6\, H\dot{\phi}G_{{3X}})
\nonumber \\
 &  & \ \ \ \ \ \ \ \,
 +18\, H(4\, H{X}^{3}G_{{4{  XXX}}}-HG_{{4}}-5\, X\dot{\phi}G_{{4\phi
X}}-G_{{4\phi}}\dot{ \phi}+7\, HG_{{4X}}X+16\, H{X}^{2}G_{{4{
XX}}}-2\,{X}^{2}\dot{\phi}G_{{4\phi{
XX}}})\nonumber \\
 &  & \ \ \ \ \ \ \ \,    +6{H}^{2}X(2\, H\dot{\phi}G_{{5{
XXX}}}{X}^{2}-6\,{X}^{2}G_{{5\phi{  XX}}}+13XH\dot{\phi}G_{{5{
XX}}}-27G_{{5\phi
X}}X+15\, H\dot{\phi}G_{{5X}}-18G_{{5\phi}})\,,\\
&&w_{4}  \equiv  2G_{4}-2XG_{5\phi}-2XG_{5X}\ddot{\phi}~.
\end{eqnarray}
For the tensor perturbations the conditions for avoidance of ghost
and Laplacian instabilities are respectively written as
\cite{DeFelice1}
\begin{equation}Q_T\equiv\frac{w_1}{4}>0\,,\, c^2_{T}\equiv\frac{w_4}{w_1}\geq 0\,.\label{csT}
\end{equation}
Conditions (\ref{csT}) are satisfied for the model under
consideration: $Q_T=1/(32\pi)>0$,  $c^2_{T}=1>0$. Taking into
account (\ref{recg2}) and (\ref{recg23X}), we rewrite functions
$c_S^2$ and $Q_S$  in the form
\begin{equation}\label{QsCub}   Q_S=-\frac{3BX^{3/2}}{4\pi a^4HX'_a\,w_{2}}=
\frac{3BX^{3/2}}{2\pi
a^4H^2X'_a}\left[-\frac{1}{2\pi}+\frac{X^{3/2}}{X'_aH^2}\left(\frac{H'_aH}{\pi
X^{1/2}}+\frac{4B}{a^4}\right)\right]^{-1}\,,\end{equation}
\begin{equation}\label{CsCub}c_S^2=\frac{1}{32\pi^2\, w_{2}^{2}\,Q_S}(w_2H-\dot w_2-4\pi w_2^2)=
\frac{1}{32\pi^2
Q_S}\left[-4\pi+H\left(\frac{a}{w_2}\right)'_a\right]\,,\end{equation}
where
\begin{equation}\label{w2} w_{2}=
\frac{H}{2}\left[\frac{1}{2\pi}-\frac{X^{3/2}}{X'_aH^2}\left(\frac{H'_aH}{\pi
X^{1/2}}+\frac{4B}{a^4}\right)\right],\, B\equiv\varepsilon
C_0\sqrt{2}\,.\end{equation}

\section{Reconstruction examples}

In this section, we will show reconstruction examples of the
gravitating scalar field theories for various laws of the Universe
expansion. Next, the theories will be tested for the absence of
pathologies in the region of applicability of cosmological models.

\subsection{Power-law expanding Universe}

Now let's consider an expanding power-law Universe:
\begin{equation}\label{pl_scal}a(t)\sim t^n \, \, \text{or}\,\,  H\sim \frac{1}{a^{3\gamma/2}}\,,\, \gamma\neq 0 \,.\end{equation}
In GR, the source of this law can be the perfect fluid:
\begin{equation}p=(\gamma-1)\rho\,, \, \rho=\frac{\rho_0}{a^{3\gamma}}\,,\end{equation}
where $w=\gamma-1$ is the EoS parameter.  Such functions
(\ref{pl_scal}) describe cosmological epochs with the radiation
($\gamma=4/3$) or matter dominated ($\gamma=1$). We define the
input functions $H(a)$, $X(a)$:
\begin{equation}\label{plHubbKin}H=\sqrt{\frac{8\pi \rho_0}{3}}\cdot\frac{1}{a^{3\gamma/2}}\,,\,
X=\frac{A^{6(\gamma-1)}}{a^{6(\gamma-1)}}\,,\, A=const>0\,, \,
\gamma\neq1\,.\end{equation} From equalities (\ref{recg2}) and
(\ref{recg23X}) we obtain
\begin{equation}\label{pl_G2}
G_2=X^{\frac{\gamma}{2(\gamma-1)}}\cdot\frac{\rho_0}{A^{3\gamma}}\left(-1+\frac{BA^{3(\gamma-1)}}{\rho_0}\right)\,,\end{equation}
\begin{equation}\label{pl_G3}G_3=X^{\frac{2-\gamma}{4(\gamma-1)}}\cdot\frac{\varepsilon}{A^{3\gamma/2}(2-\gamma)}
\sqrt{\frac{\rho_0}{12\pi}}\cdot\left(\gamma-\frac{BA^{3(\gamma-1)}}{\rho_0}\right)+const\,,
\, \gamma\neq2\,.\end{equation}

Conditions for absence of ghosts and Laplacian instabilities
\begin{equation}\label{}c^2_S=\frac{\rho_0}{3BA^{3\gamma-3}}\left(\gamma(3\gamma-2)-
\frac{BA^{3\gamma-3}}{\rho_0}\right)\geq 0\,,\end{equation}
\begin{equation}\label{}Q_S=\frac{3A^{3\gamma-3}B}{8\pi\rho_0\left(\gamma-2+\frac{BA^{3\gamma-3}}{\rho_0}\right)}>0
\end{equation} are satisfied in two cases. Firstly, for $B>0$\,: \begin{equation}\label{BA}2-\gamma<\frac{BA^{3\gamma-3}}{\rho_0}\leq\gamma(3\gamma-2)\,.\end{equation}
The necessary condition for the fulfillment of (\ref{BA}) has the
form: $\omega<-5/3$, $\omega>0$. Secondly, for $B<0$\,:
\begin{equation}\label{AB}\gamma-2<\frac{|B|A^{3\gamma-3}}{\rho_0}\leq\gamma(2-3\gamma)\,.\end{equation}
The necessary condition for the fulfillment of (\ref{AB}) has the
form: $-1<\omega<-1/3$. This model will not describe pressureless
dark matter ($\gamma= 1$). Below we present a model with this
substance. We observe that the conditions for absence of ghosts
and Laplacian instabilities are always satisfied ($a(t)>0$), and
thus the scenario at hand is free from pathologies of the scalar
sector at both background and perturbative level. Thus, the KGB
theory can describe power-law expanding Universes by means of a
nonzero scalar charge ($B\neq 0$) and without other matter.

\subsection{Exponential-law expanding Universe}

\begin{figure}[h]
\includegraphics[width=7cm]{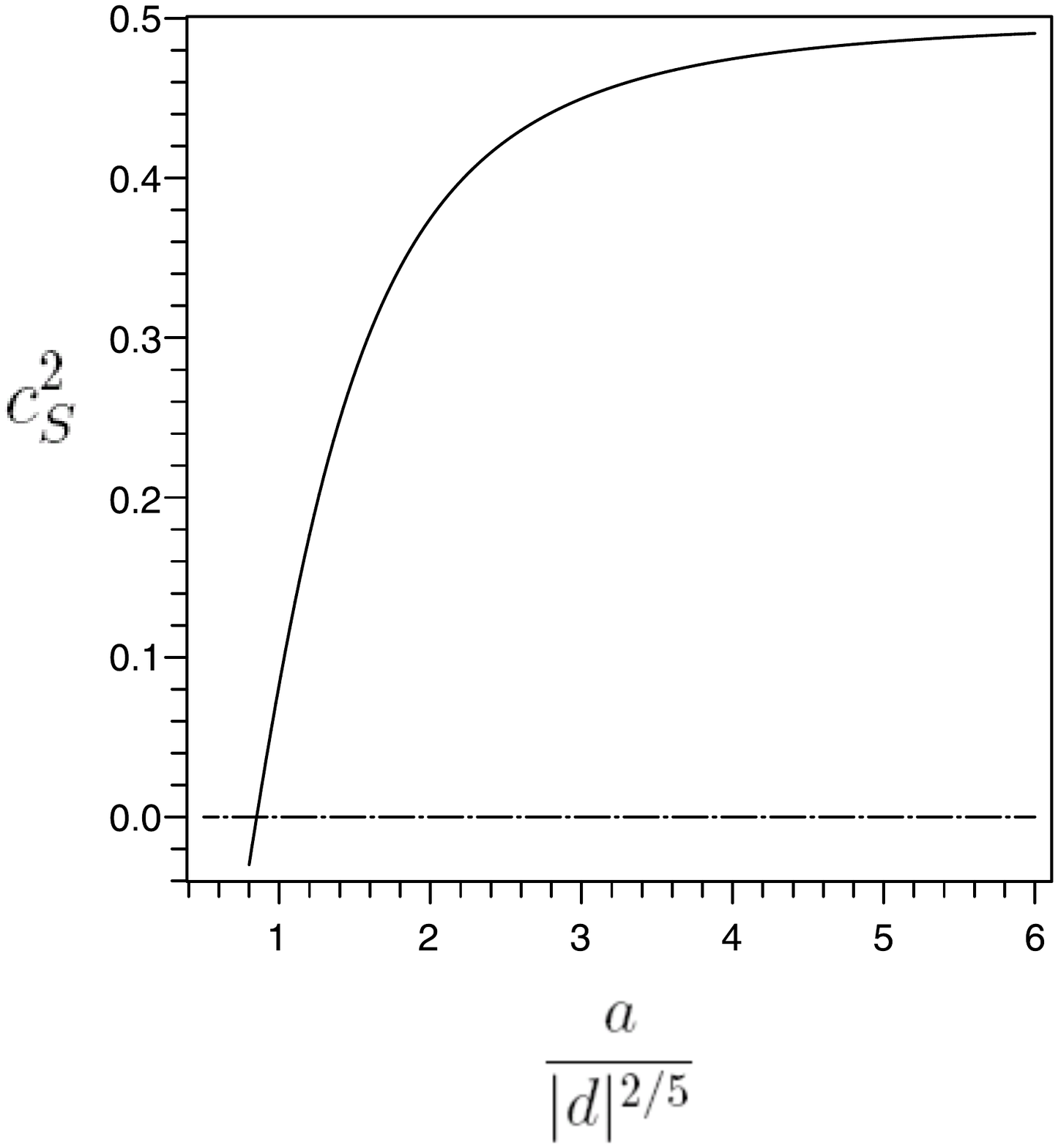}
\caption{Profile of the sound speed squared $c_S^2$.
\label{Infcs}}
\end{figure}

Here we will consider a dark energy equation of state $p=-\rho$\,.
The Hubble parameter has the constant value
\begin{equation}\label{ExpHabbl}H=\sqrt{\frac{8\pi \rho_0}{3}}=const\,, \, a=a_i\exp(Ht)\,.\end{equation}
We take the kinetic density $X(a)$ in the form
\begin{equation}\label{ExpKinetic}X=A^2a\,, \, A=const>0\,.\end{equation}
The two conditions (\ref{QsCub}), (\ref{CsCub}) for absence of
pathologies in the scalar perturbations become
\begin{equation}\label{InfPer}Q_S=\frac{3}{8\pi}\cdot\frac{|d|}{|d|+a^{5/2}}>0\,,\,
c^2_S=\frac{\frac{a^{5/2}}{2}-\frac{|d|}{3}}{|d|+a^{5/2}}\geq0\,,\end{equation}
where it is supposed $d\equiv\dfrac{3AB}{\rho_0}<0$ ($B<0$). The
condition $Q_S>0$ is satisfied for any values $a\geq0$. The
inequality $c^2_S>0$ holds for $a\geq
a_*=\left(\dfrac{2|d|}{3}\right)^{2/5}$. Assuming that the gravity
theory without quantum consideration works since the Planck times
$t\gtrsim t_p$, it is natural to put $a_*=a_p=a(t_p)$. As the
constant $\rho_0$, take the Planck density $\rho_0=\rho_p$. The
profile of $c^2_S$ is shown in Fig. (\ref{Infcs}).

From equalities (\ref{recg2}) and (\ref{recg23X}) we obtain
\begin{equation}\label{ExpG23}G_2=-\rho_p\left(1+\frac{|d|A^5}{3X^{5/2}}\right)\,,\,
G_3=\frac{\varepsilon
|d|A^5\sqrt{\rho_p}}{12\sqrt{3\pi}X^3}+const\,.\end{equation} The
proposed model can describe the early inflation of the Universe
since the Planck time without pathologies at the perturbative
level (\ref{InfPer}). The area with pathology is "hidden" behind
the Planck boundary $t=t_p$, where the gravity theory without
quantum consideration has a dubious status.

\subsection{Unification of dark matter - dark energy}

In $\Lambda$CDM paradigm, the description of the transition to
late the Universe acceleration is obtained by the usual
consideration of a pressureless dark matter (DM) with the EoS
parameter $w_m=0$ and a cosmological constant with the EoS
parameter $w_{\Lambda}=-1$:
\begin{equation}\label{qaq}\frac{3H^2}{8\pi}=\rho=\rho_{de}+\frac{\rho_{dm}}{a^3}\,; \, \,\rho_{de}\,, \rho_{dm}=const\,,\end{equation}
where $a(t_0)=1$, $t_0$ -- the present moment of time. The
deceleration parameter (DP)
\begin{equation}\label{trdp}q(a)=\frac{d}{dt}\left(\frac{1}{H}\right)-1=\frac{a^3_{*}-a^3}{2a^3_{*}+a^3}\end{equation}
changes sign. There are two phases: $0 \leq a \leq a_*$ and
$a>a_*$, where
\begin{equation}\label{q0}q(a_*)=0\,, \, a_*=\left(\frac{\rho_{dm}}{2\rho_{de}}\right)^{1/3}=
\left(\frac{\Omega_{dm}}{2\Omega_{de}}\right)^{1/3}\,.\end{equation}
Here, $\Omega_{dm}$, $\Omega_{de}$ -- the dimensionless density
parameters. In the first phase there is no acceleration ($q\geq
0$), and the second one is characterized by the Universe
acceleration ($q<0$).

\begin{figure}[h]
\includegraphics[width=7cm]{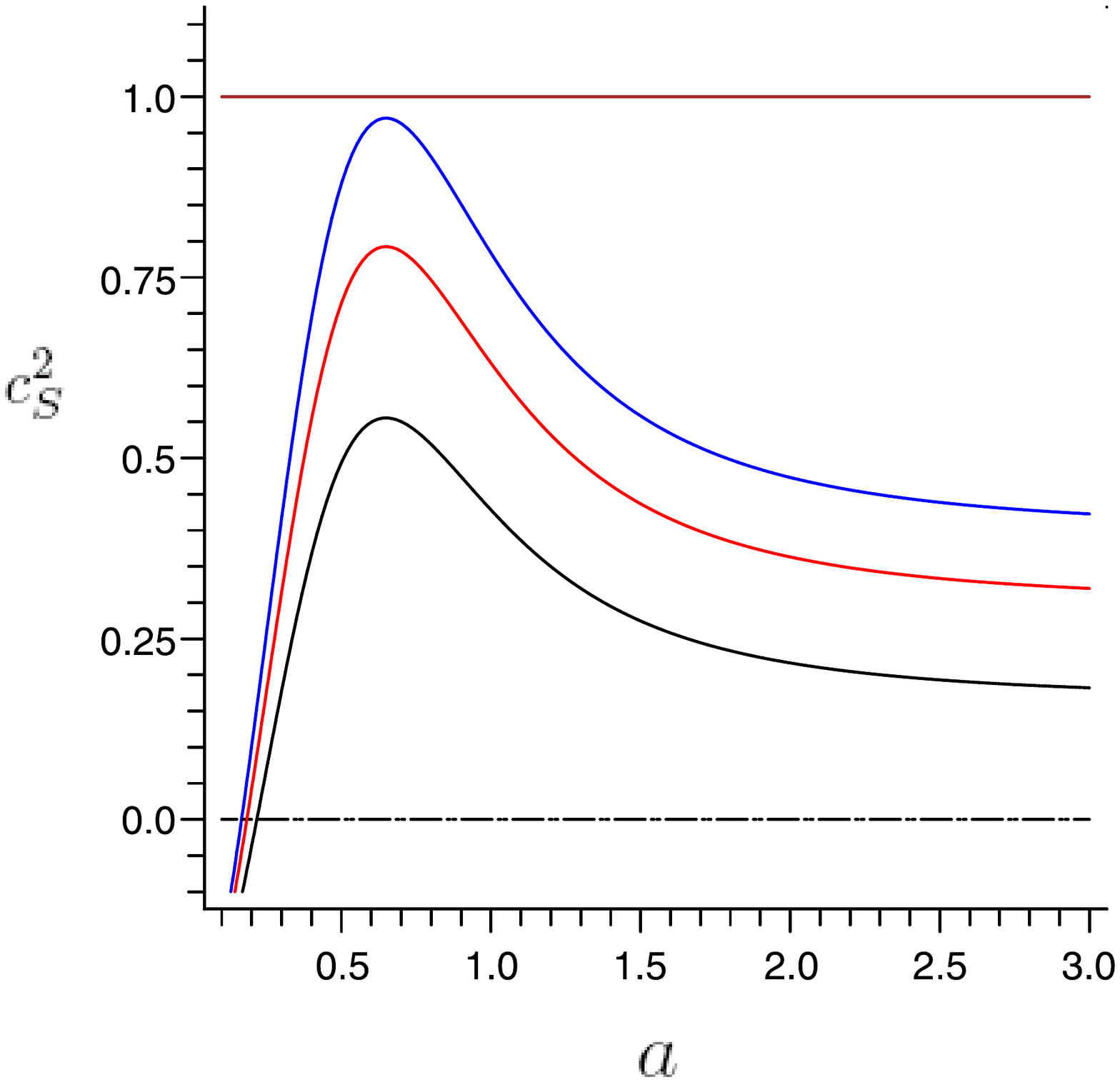}
\caption{Profiles of the sound speeds squared $c_S^2$ for
$a_*=0.555$ and $\alpha=0.75$; $0.95$; $1.1$ (black, red, blue).
The value $a_*$ corresponds a inflexion point ($q=0$). The sound
speed squared is positive, starting with small values of the scale
factor $a$. \label{cs2MatterInf}}
\end{figure}

We will present a specific model, which presents a unified
description of the DM and DE sectors. We define the input
functions $H(a)$, $X(a)$:
\begin{equation}\label{trHubbKin}H=\sqrt{\frac{8\pi}{3}\left(\rho_{de}+\frac{\rho_{dm}}{a^3}\right)}\,,\,
X=\rho_{de}+\frac{\rho_{dm}}{a^3}\,.\end{equation} From equalities
(\ref{recg2}) and (\ref{recg23X}) we obtain
\begin{equation}\label{tr_G2}
G_2=-X-\frac{\rho_{de}BX^{1/2}}{\rho_{dm}}+\frac{BX^{3/2}}{\rho_{dm}}\,,\end{equation}
\begin{equation}\label{tr_G3}G_{3}=\frac{\varepsilon}{4\rho_{dm}\sqrt{3\pi}}\left(\rho_{dm}\ln \frac{X}{X_0}-2BX^{1/2}\right)+const\,.\end{equation}

The two conditions (\ref{QsCub}), (\ref{CsCub}) for absence of
pathologies in the scalar perturbations become
\begin{equation}\label{trQsCub}Q_S=\frac{3}{8\pi}>0\,,\end{equation}
\begin{equation}\label{trCsCub}c^2_S=\frac{\rho_{dm}\left(\rho_{de}+\frac{4\rho_{dm}}{a^3}\right)
}{3B\left(\rho_{de}+\frac{\rho_{dm}}{a^3}\right)^{3/2}}-\frac{1}{3}=
\frac{2\alpha}{3}\frac{1+\frac{8a_*^3}{a^3}}{\left(1+\frac{2a_*^3}{a^3}\right)^{3/2}}-\frac{1}{3}\geq0\,,\end{equation}
where $\alpha=\rho_{de}^{1/2}\cdot a^3_*/B$. It is assumed that
$B>0$. There are values of the parameter $\alpha$ for which,
starting from some value of the scale factor $a$, the stability
condition (\ref{trCsCub}) is fulfilled. The stability conditions
are violated at small values of $a$. However, for $a\sim 0$, model
(\ref{qaq}) is not valid, and therefore the question of stability
at small $a$ is not important.  The profile of $c^2_S$ is shown in
Fig. \ref{cs2MatterInf}.

Thus, the KGB theory can describe the late transitional Universe
by means of a nonzero scalar charge ($B\neq 0$) and without other
matter.

\subsection{Post-inflationary transition to the radiation-dominated
phase}

A fluid with the equation of state
\begin{equation}\label{eosInflRad}p=\frac{\rho}{3}-\mu\rho^2\,,\, \mu>0\end{equation} describes the
transition from the inflationary era to the radiation era in the
the early Universe \cite{Pierre}. The linear equation of state
$p=\rho/3$ corresponds to radiation. The polytropic equation of
state $p=-\mu\rho^2$ may be due to Bose-Einstein condensates with
repulsive ($\mu < 0$) or attractive ($\mu > 0$) self-interaction,
or have another origin. We will construct a scalar analogue of the
theory (\ref{eosInflRad}).

We write the DP as follows
\begin{equation}\label{InflRaddp}q(\rho)=\frac{\rho+3p}{2\rho}=1-\frac{3\mu\rho}{2}=1-\frac{\rho}{\rho_d}\,,\end{equation}
where the parameter $\rho_d$ corresponds a inflexion point:
\begin{equation}  q(\rho_d)=0\,,\, \rho_d=\frac{2}{3\mu}\,.\end{equation}
The Universe is accelerating when $\rho>\rho_d$ and decelerating
when $\rho\leq\rho_d$. The "equation of continuity"
\begin{equation}\dot\rho+3H(\rho+\rho)=0\end{equation}
can be integrated into $\rho=\dfrac{2\rho_d}{a^4+1}$, where we
assume that $a=a_d=1$ corresponds to the end of inflation. When
$a\ll 1$, the density $\rho$ takes the maximum value $2\rho_d$.
The Planck density $\rho_p$  will be determined so that it
approximately corresponds to he maximum value: $\rho_p\approx
2\rho_d$. Thus the density will take the form
\begin{equation}\rho=\frac{\rho_p}{a^4+1}\,,\end{equation}
which gives the Hubble parameter
\begin{equation}\label{InflRadHabl}H^2=\frac{8\pi \rho_p}{3}\cdot\frac{1}{a^4+1}\,.\end{equation}
During the inflationary phase ($a\ll 1$), the parameter $H$ has an
approximately constant value $H\simeq\sqrt{\dfrac{8\pi
\rho_p}{3}}$\,. After the inflation ($a\gg 1$), the Universe
enters in the radiation era: $H\simeq\sqrt{\dfrac{8\pi
\rho_p}{3a^4}}$\,. This model was considered in the work
\cite{Pierre}.

\begin{figure}[h]
\includegraphics[width=7cm]{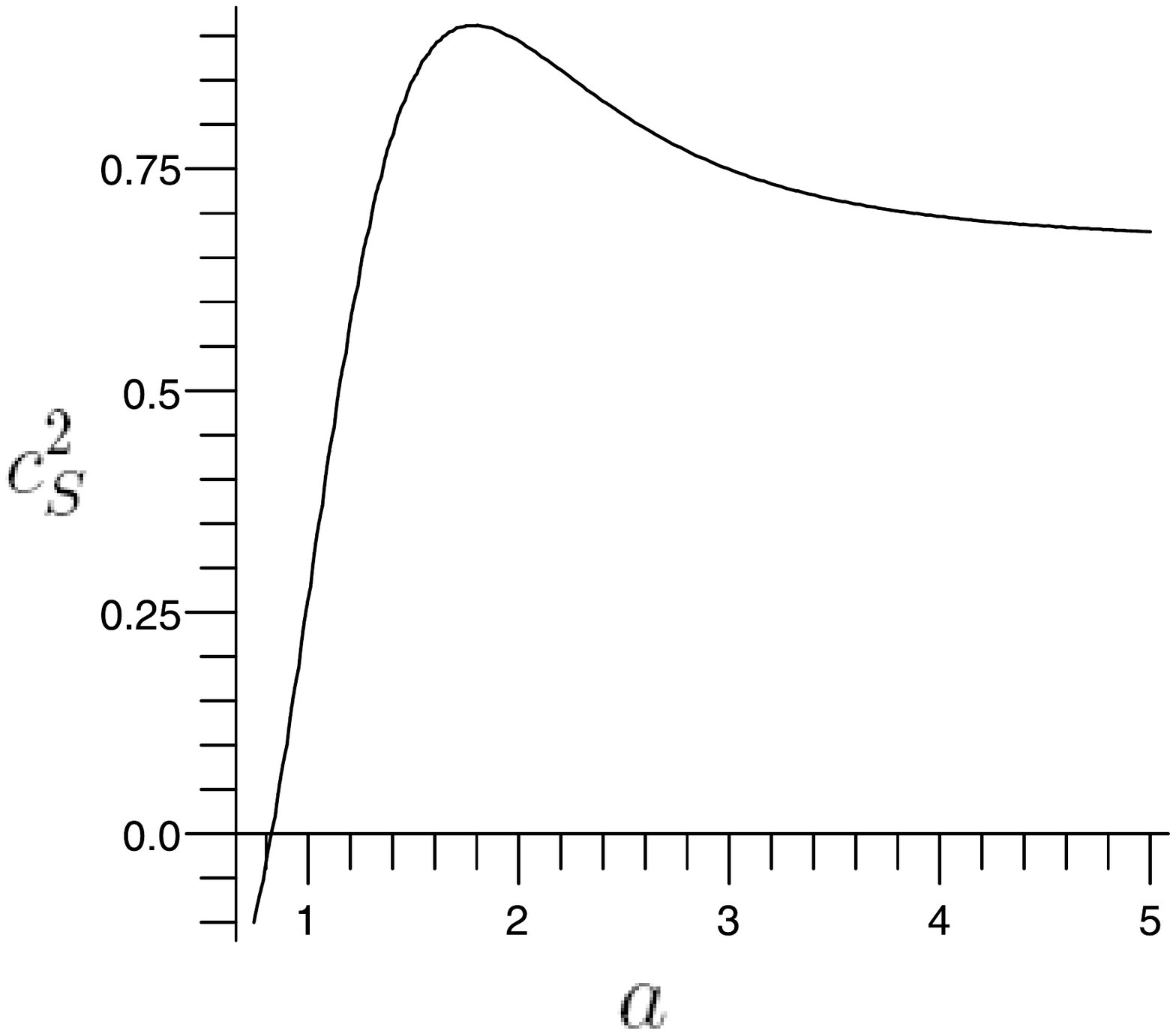}
\caption{Profile of the sound speed squared $c_S^2$ for $d=8/3$.
The value $a=a_d=1$ corresponds a inflexion point ($q=0$). The
sound speed squared is  positive for $a\gtrsim 0.82815$.
\label{InfRad}}
\end{figure}

We chose the Hubble parameter $H$ according to the task. We take
the kinetic density $X(a)$ in the form
\begin{equation}\label{InflRadKinetic} X(a)=\frac{A^2}{a^2}\,,\, A=const>0\,.\end{equation}
The two conditions (\ref{QsCub}), (\ref{CsCub}) for absence of
pathologies in the scalar perturbations become
$$Q_S=\frac{3d\cdot(1+a^4)^2}{8\pi[2a^4(1-a^4)+d\cdot(1+a^4)^2]}>0\,,$$
\begin{equation}\label{InflRadLaplac} c^2_S=\frac{-d^2\cdot(1+a^4)^4+2d\cdot
a^4(1+a^4)^2(5a^4+3)+16a^{12}(3-a^4)}{3d\cdot(1+a^4)^2[2a^4(1-a^4)+d\cdot(1+a^4)^2]}
\geq 0\,,\end{equation} where $d\equiv\dfrac{3AB}{\rho_p}$\,. They
have the following limits:
\begin{equation}\frac{3}{8\pi}\leftarrow Q_S\rightarrow\frac{3d}{8\pi (d-2)}\,\, \, \text{as} \,\,\, 0\leftarrow a\rightarrow +\infty\,,\end{equation}
\begin{equation}-\frac{1}{3}\leftarrow c^2_S \rightarrow\frac{8-d}{3d}\,\, \, \text{as} \,\,\,  0\leftarrow a\rightarrow +\infty \,, \,d\neq 2\,.\end{equation}
In case $d>2$, the condition $Q_S>0$ is satisfied for any values
$a\geq0$. There are values of the parameter $d$ for which,
starting from some value of the scale factor $a$, the stability
condition (\ref{InflRadLaplac}) is fulfilled.  The Fig.
\ref{InfRad} shows that the sound speeds squared become negative
at early times, $a\lesssim 0.82815$ (for $d=8/3$). The transition
point $a_d=1$ to unaccelerated expansion of the Universe is in the
region of stability of the model for $d=8/3$.

From equalities (\ref{recg2}) and (\ref{recg23X}) we obtain
$$G_2=\frac{\rho_p\cdot X^2(d\cdot X^2+(d-3)A^4)}{3A^4(A^4+X^2)}\,,
$$
\begin{equation}G_3=\frac{\varepsilon}{2}\sqrt{\frac{\rho_p}{3\pi}}\int^XdX\cdot
\left[\frac{A^4}{X^{1/2}(A^4+X^2)^{3/2}}-\frac{d\cdot(A^4+X^2)^{1/2}}{4A^4X^{1/2}}\right]+const\,.\end{equation}
The model is unstable for $a\ll 1$. During  the radiation era
($a\gg1$), the functions $G_2$, $G_3$ have approximately view:
\begin{equation}
G_2\simeq\frac{\rho_p(d-3)\cdot X^2}{3A^4}\,,\,
G_3\simeq\frac{\varepsilon(4-d)}{4A^2}\cdot\sqrt{\frac{\rho_p\cdot
X}{3\pi }}+const\,.\end{equation}

\section{Conclusion}

The HG theory  action density (\ref{lagr3}) contains arbitrary
functions $G_i(X,\phi)$, $i=2,3,4,5$. The standard algorithm
assumes setting the functions $G_i(X,\phi)$, obtaining a solution
for the scale factor $a(t)$ and the scalar field $\phi(t)$, and
analyzing them for compliance with the observed data. In HG
theory, as a rule, this approach requires considerable effort and
the involvement of numerical methods for solving a differential
equations system. The search  result for a solution is not
predictable. Another way is the reconstruction method. The
evolution scenario of the Universe is set {\it a priori}. It
serves to limit and determine the parameters of the studied
modified  gravity theory. In this article, we have proposed a way
to reconstruct the KGB theory with shiftsymmetric based on the
Hubble evolution in the flat FRW spacetimes. The dynamic solution
$H(a)$ is provided by a nonzero scalar charge associated with the
shift symmetry. In the described method, the choice of the Hubble
parameter $H(a)$ does not uniquely determine the scalar theory. To
complete the construction, we need to set the kinetic density
$X(a)$. This freedom is used to eliminate the instability of
perturbations in the range of applicability of the considered
cosmological models. The process of selecting function $X(a)$  is
heuristic.  We have illustrated this method for models: the
perfect fluid, the unified description dark energy-dark matter,
the post-inflationary transition to the radiation-dominated phase.

We have succeeded in constructing a model  of  perfect fluid
($\omega\neq-1$) that is stable throughout the expansion history.
Other models contain the instability on a set of values $a\in [0,
a_1)$. This set is not included in the range of applicability of
cosmological models (\ref{ExpHabbl}) and (\ref{qaq}), therefore
the question of stability is not important on $[0, a_1)$ for these
models. The post-inflationary transition model (\ref{eosInflRad})
is stable around the transition point $a_d$ and in the future.

\acknowledgments This work is supported by the Russian Foundation
for Basic Research (Grant No. 19-52-15008).

\end{document}